\documentclass[preprint,showpacs,preprintnumbers]{revtex4}
\usepackage{amssymb}
\usepackage{amsmath}
\usepackage{graphicx}
\usepackage{dcolumn}
\usepackage{bm}

\newcommand{\be}{\begin{equation}}
\newcommand{\ee}{\end{equation}}
\newcommand{\bea}{\begin{eqnarray}}
\newcommand{\eea}{\end{eqnarray}}

\input{tcilatex}
\begin{document}

\title{Constructing quantum games from symmetric non-factorizable joint
probabilities}
\author{James M.~Chappell}
\email{james.m.chappell@adelaide.edu.au}
\affiliation{School of Chemistry and Physics, University of Adelaide, South Australia
5005, Australia }
\author{Azhar Iqbal}
\affiliation{a) School of Electrical and Electronic Engineering, University of Adelaide,
South Australia 5005, Australia\\
b) Centre for Advanced Mathematics and Physics, National University of
Sciences \& Technology, Peshawar Road, Rawalpindi, Pakistan}
\author{Derek Abbott}
\affiliation{School of Electrical and Electronic Engineering, University of Adelaide
5005, Australia }
\date{\today}

\begin{abstract}
We construct quantum games from a table of non-factorizable joint
probabilities, coupled with a symmetry constraint, requiring symmetrical
payoffs between the players. We give the general result for a Nash
equilibrium and payoff relations for a game based on non-factorizable joint
probabilities, which embeds the classical game. We study a quantum version
of Prisoners' Dilemma, Stag Hunt, and the Chicken game constructed from a
given table of non-factorizable joint probabilities to find new outcomes in
these games. We show that this approach provides a general framework for
both classical and quantum games without recourse to the formalism of
quantum mechanics.
\end{abstract}

\maketitle

\section{Introduction}

As an established branch of mathematics, game theory \cite{Binmore,Rasmusen}
analyzes strategic decision making of competing agents who find themselves
in conflict situations. Game theory finds extensive applications in social
sciences, biology and engineering. Recent developments in quantum computing
and quantum information theory \cite{Nielsen} have motivated efforts to
recast classical game theory using quantum probability amplitudes \cite%
{Peres}, leading to the birth of the area of quantum games \cite%
{Mermin,Mermin1,MeyerDavid,EWL,Vaidman,Brandt,BenjaminHayden1,BenjaminHayden2,EnkPike,Johnson,Johnson1,MarinattoWeber,IqbalToor1,DuLi,Du,Piotrowski,Piotrowski1,FlitneyAbbott1,Iqbal,FlitneyAbbott2,CheonTsutsui,Shimamura,IchikawaTsutsui,IqbalCheonM,IqbalWeigert,Ozdemir,IqbalCheon2007,FlitneyHollenberg1,IqbalCheonAbbott,Ichikawa,IqbalAbbott2008,Chappell,FlitneyRecent}
. Quantum games have been proposed \cite{MeyerDavid} as a new way to
approach quantum algorithms, including speculation that nature may be
playing quantum games at the molecular level \cite{Dawkins}.

In the area of quantum games, a recently reported \cite{IqbalCheon2007}
probabilistic approach constructs them from a set of non-factorizable joint
probabilities. It is a set that cannot be factorized in terms of other more
fundamental probabilities. The motivating idea being that Bell's
inequalities can be violated by a set of non-factorizable probabilities,
though this does not imply that any non-factorizable set will violate the
inequalities. As the violation of Bell's inequalities is considered a
fundamentally quantum aspect, one is motivated to have an approach to
quantum games that constructs them from the property of a probability set of
being non-factorizable. This then provides greater mathematical generality
for quantum games, providing accessibility without the formalism of quantum
mechanics.

The earliest quantization attempts \cite{EWL} were focussed at the
two-player two-strategy ($2\times 2$) non-cooperative games . The playing of
a $2\times 2$ game can be easily realized if players share a physical system
that involves $4$ joint probabilities. This, for instance, will be the case
when players share two coins that can be put in head ($\mathcal{H}$) and
tail ($\mathcal{T}$) states. A referee gives each player a coin and asks
them to flip it to either the $\mathcal{H}$ or the $\mathcal{T}$ state. As
no tossing but only flipping actions are involved, it does not matter if the
coins are biased or not. After a players' flipping (or not-flipping) actions
the coins can be found in a $\mathcal{HH}$, $\mathcal{HT}$, $\mathcal{TH}$,
or $\mathcal{TT}$ state, where the first entry in each pair, is reserved for
the state of Alice's coin. From a given $2\times 2$ game table, the referee
can then award players their payoffs depending on the state of the two coins.

The probabilistic approach to quantum games developed in Ref. \cite%
{IqbalCheon2007} extends the playing of a $2\times 2$ game towards the
quantum domain by considering two players who share a bigger physical system
that also involves coin tossing. Consider four biased coins that two players
share to play a $2\times 2$ game according to the following arrangement. In
a run, each player is given two coins and she/he has to select one. The
referee tosses the two selected coins together and records the outcome. It
can then be shown that the players' payoffs, in a mixed-strategy version of
the $2\times 2$ game, can be expressed in terms of their strategic choices
performed over multiple runs and the relevant $16$ factorizable joint
probabilities.

In order to maintain the bilinear payoff structure of the $2\times 2$ game,
constraints are placed on joint probabilities. Allowing joint probabilities
to become non-factorizable, while they remain under these constraints,
ensures that the classical payoffs and the outcome of the game are obtained
when the probabilities become factorizable.

In the present paper, we introduce an extra condition requiring that the
considered joint probabilities are also symmetric both when they are
factorizable and non-factorizable. We argue that this is a natural
constraint to be included in a probabilistic approach to quantum games that
builds them from non-factorizable joint probabilities. We find that this
constraint further narrows down our focus in obtaining a quantum game from
probabilistic considerations only. We study quantum versions of the
Prisoners' Dilemma (PD) game, the Stag Hunt (SH) game, and the Chicken game 
\cite{Binmore,Rasmusen} within this approach that constructs them from not
only non-factorizable but also symmetric joint probabilities. We investigate
how non-factorizable and symmetric joint probabilities permit new equilibria
in these games. By introducing parameters that present a measure of
non-factorizability, we discuss a novel way of obtaining a set of
non-factorizable joint probabilities, which allows us to identify
factorizable, non-factorizable, and the quantum domains.

\section{An approach towards extending a $2\times 2$ game}

We begin with the table of sixteen joint probabilities as shown in Table \ref%
{RawEPRprobabilities}.

\begin{table}[h]
\begin{equation}
\begin{array}{c}
\text{Alice}%
\end{array}%
\begin{array}{c}
\underset{}{%
\begin{array}{c}
S_{1}%
\end{array}%
\begin{array}{c}
+1 \\ 
-1%
\end{array}%
} \\ 
\overset{}{%
\begin{array}{c}
S_{2}%
\end{array}%
\begin{array}{c}
+1 \\ 
-1%
\end{array}%
}%
\end{array}%
\overset{\overset{%
\begin{array}{c}
\text{Bob}%
\end{array}%
}{%
\begin{array}{cc}
\overset{%
\begin{array}{c}
S_{1}^{\prime }%
\end{array}%
}{%
\begin{array}{cc}
+1 & -1%
\end{array}%
} & \overset{%
\begin{array}{c}
S_{2}^{\prime }%
\end{array}%
}{%
\begin{array}{cc}
+1 & -1%
\end{array}%
}%
\end{array}%
}}{%
\begin{tabular}{c|c}
$\underset{}{%
\begin{tabular}{ll}
$p_{1}$ & $p_{2}$ \\ 
$p_{3}$ & $p_{4}$%
\end{tabular}%
}$ & $\underset{}{%
\begin{tabular}{ll}
$p_{5}$ & $p_{6}$ \\ 
$p_{7}$ & $p_{8}$%
\end{tabular}%
}$ \\ \hline
$\overset{}{%
\begin{tabular}{ll}
$p_{9}$ & $p_{10}$ \\ 
$p_{11}$ & $p_{12}$%
\end{tabular}%
}$ & $\overset{}{%
\begin{tabular}{ll}
$p_{13}$ & $p_{14}$ \\ 
$p_{15}$ & $p_{16}$%
\end{tabular}%
}$%
\end{tabular}%
}
\end{equation}%
\caption{General probability table.}
\label{RawEPRprobabilities}
\end{table}
For Alice and Bob we also have payoff matrices given by 
\begin{equation}
\mathcal{A}=%
\begin{array}{c}
\text{Alice}%
\end{array}%
\begin{array}{c}
S_{1} \\ 
S_{2}%
\end{array}%
\overset{\overset{%
\begin{array}{c}
\text{Bob}%
\end{array}%
}{%
\begin{array}{cc}
S_{1}^{\prime } & S_{2}^{\prime }%
\end{array}%
}}{\left( 
\begin{array}{cc}
a_{1} & a_{2} \\ 
a_{3} & a_{4}%
\end{array}%
\right) },\text{ \ \ }\mathcal{B}=%
\begin{array}{c}
\text{Alice}%
\end{array}%
\begin{array}{c}
S_{1} \\ 
S_{2}%
\end{array}%
\overset{\overset{%
\begin{array}{c}
\text{Bob}%
\end{array}%
}{%
\begin{array}{cc}
S_{1}^{\prime } & S_{2}^{\prime }%
\end{array}%
}}{\left( 
\begin{array}{cc}
b_{1} & b_{2} \\ 
b_{3} & b_{4}%
\end{array}%
\right) },  \label{A and B matrices}
\end{equation}%
giving Alice's and Bob's payoffs, respectively. Because we are considering
games with symmetrical payoffs we have $\mathcal{B}=\mathcal{A}^{T}$, where $%
T$ indicates transpose. This requires

\begin{equation}
b_{1}=a_{1},\text{ }b_{2}=a_{3},\text{ }b_{3}=a_{2},\text{ }b_{4}=a_{4}.
\end{equation}%
In a mixed-strategy game one has the strategy vectors $\mathbf{x=(}%
x,1-x)^{T} $ and $\mathbf{y=(}y\mathbf{,}1-y\mathbf{)}^{T}$, with $x,y$ $\in
\lbrack 0,1]$ giving the probabilities for Alice and Bob to choose $S_{1}$
and $S_{1}^{\prime }$ respectively. Payoff relations in a mixed-strategy
game are

\begin{equation}
\Pi _{A,B}(x,y)=\mathbf{x}^{T}\mathcal{(A},\mathcal{B)}\mathbf{y,}
\label{mixed-strategy payoffs}
\end{equation}%
where subscripts $A$ and $B$ refer to Alice and Bob, respectively. In this
notation we can, for instance, have the pure strategy payoffs $\Pi
_{A}(S_{1},S_{1}^{\prime })=\Pi _{A}(1,1)=a_{1}=\Pi _{B}(1,1)$ etc. We
represent strategies $\mathbf{x}$ and $\mathbf{y}$ by numbers $x$ and $y$
respectively and note that the strategy pair $(x^{\star },y^{\star })$ is a
Nash equilibrium (NE) when

\begin{equation}
\Pi _{A}(x^{\star },y^{\star })-\Pi _{A}(x,y^{\star })\geq 0,\text{ \ \ }\Pi
_{B}(x^{\star },y^{\star })-\Pi _{B}(x^{\star },y)\geq 0.  \label{NE}
\end{equation}

A possible physical realization for playing this symmetric game uses two
coins in the following arrangement. The referee announces the association $%
S_{1},S_{1}^{\prime }\sim \mathcal{H}$ and $S_{2},S_{2}^{\prime }\sim 
\mathcal{T}$ and each player's strategy consists of secretly flipping
his/her penny either to the $\mathcal{H}$ or to the $\mathcal{T}$ state. The
players then simultaneously return their pennies to the referee. The referee
observes the state of the two coins and rewards the players. In the case of
pure strategies the referee can use the matrices (\ref{A and B matrices})
and in case of mixed strategies, the players are rewarded according to the
payoff relations (\ref{mixed-strategy payoffs}).

As mentioned in the introduction, the referee can also have a different
arrangement that allows the playing the game using four coins instead of
two, as follows. S/he identifies the four coins as $S_{1},S_{2};S_{1}^{%
\prime },S_{2}^{\prime }$ (note that $S_{1}$ and $S_{2}$ are no longer a
player's actions to put his/her penny in $\mathcal{H}$ or $\mathcal{T}$
state). In a run s/he gives coins $S_{1},S_{2}$ to Alice and coins $%
S_{1}^{\prime },S_{2}^{\prime }$ to Bob. Each player now has to choose one
out of the two coins so that the chosen pair is one of the $%
(S_{1},S_{1}^{\prime }),$ $(S_{1},S_{2}^{\prime }),$ $(S_{2},S_{1}^{\prime
}),$ $(S_{2},S_{2}^{\prime }) $. The players return the two chosen coins to
the referee who tosses them together and records the outcome. The referee
then collects four coins (two tossed and two untossed) and prepares them for
the next run. In this extended game, players' payoff relations can now be
defined by making the association $\mathcal{H}\thicksim +1$ \& $\mathcal{T}%
\thicksim -1$ and using the $16$ joint probabilities $p_{1},p_{2},...p_{16}$%
, as described in Table (\ref{RawEPRprobabilities}).

We write the payoffs relations as,

\begin{eqnarray}
\Pi _{A,B}(S_{1},S_{1}^{\prime }) &=&\tsum\nolimits_{i=1}^{4}(a,b)_{i}p_{i},%
\text{ \ \ \ \ \ }\Pi _{A,B}(S_{1},S_{2}^{\prime
})=\tsum\nolimits_{i=5}^{8}(a,b)_{i-4}p_{i},  \notag \\
\Pi _{A,B}(S_{2},S_{1}^{\prime })
&=&\tsum\nolimits_{i=9}^{12}(a,b)_{i-8}p_{i},\text{ \ \ }\Pi
_{A,B}(S_{2},S_{2}^{\prime })=\tsum\nolimits_{i=13}^{16}(a,b)_{i-12}p_{i},
\label{StrategyPayoffs}
\end{eqnarray}%
where $\Pi _{B}(S_{2},S_{1}^{\prime })$, for instance, corresponds when
Alice selects her $S_{2}$ coin and Bob selects his $S_{1}^{\prime }$ coin
over all the runs. In Eqs.~(\ref{StrategyPayoffs}) each of the four payoff
relations give mixed-strategy payoffs of the $2\times 2$ game. Over many
runs, the players can also select a probability distribution over the
available strategies and one can define

\begin{equation}
\Pi _{A,B}(x,y)=\mathbf{x}^{T}%
\begin{pmatrix}
\Pi _{A,B}(S_{1},S_{1}^{\prime }) & \Pi _{A,B}(S_{1},S_{2}^{\prime }) \\ 
\Pi _{A,B}(S_{2},S_{1}^{\prime }) & \Pi _{A,B}(S_{2},S_{2}^{\prime })%
\end{pmatrix}%
\mathbf{y},  \label{MixedStrategyPayoffsExtendedGame}
\end{equation}%
where $x\in \lbrack 0,1]$ is the probability with which Alice selects coins $%
S_{1}$ and $y\in \lbrack 0,1]$ is the probability with which Bob selects
coin $S_{1}^{\prime }$. Eqs.~(\ref{MixedStrategyPayoffsExtendedGame}) give
mixed-strategy payoffs in an extension of the original $2\times 2$ game. As
discussed in Refs. \cite{IqbalCheon2007,IqbalCheonAbbott,IqbalAbbott2008},
this extension involving $16$ joint factorizable probabilities can be
considered a re-expression of the classical game that transforms the
original game in such a way that a transition to the quantum game is
achievable by a consideration of non-factorizable joint probabilities. This
re-expressed game with factorizable probabilities is of course classically
implementable and is not to be confused with the original game from which it
is derived.

\subsection{Obtaining a symmetric game}

Enforcing game symmetry, under the interchange of players, we require $\Pi
_{A}(S_{i},S_{j}^{\prime })=\Pi _{B}(S_{j},S_{i}^{\prime })$ and thus we
note from Eqs.~(\ref{StrategyPayoffs}) that this is achievable if 
\begin{equation}
p_{2}=p_{3};\text{ \ \ }p_{5}=p_{9},\,p_{6}=p_{11},\,p_{7}=p_{10},%
\,p_{8}=p_{12};\text{ \ \ }p_{14}=p_{15}.
\end{equation}%
With this we produce Table (\ref{SymmTableEPRprobabilities}) and notice that
the table of probabilities is now symmetric across the main diagonal. 
\begin{table}[h]
\begin{equation}
\begin{array}{c}
\text{Alice}%
\end{array}%
\begin{array}{c}
\underset{}{%
\begin{array}{c}
S_{1}%
\end{array}%
\begin{array}{c}
+1 \\ 
-1%
\end{array}%
} \\ 
\overset{}{%
\begin{array}{c}
S_{2}%
\end{array}%
\begin{array}{c}
+1 \\ 
-1%
\end{array}%
}%
\end{array}%
\overset{\overset{%
\begin{array}{c}
\text{Bob}%
\end{array}%
}{%
\begin{array}{cc}
\overset{%
\begin{array}{c}
S_{1}^{\prime }%
\end{array}%
}{%
\begin{array}{cc}
+1 & -1%
\end{array}%
} & \overset{%
\begin{array}{c}
S_{2}^{\prime }%
\end{array}%
}{%
\begin{array}{cc}
+1 & -1%
\end{array}%
}%
\end{array}%
}}{%
\begin{tabular}{c|c}
$\underset{}{%
\begin{tabular}{ll}
$p_{1}$ & $p_{2}$ \\ 
$p_{2}$ & $p_{4}$%
\end{tabular}%
}$ & $\underset{}{%
\begin{tabular}{ll}
$p_{5}$ & $p_{6}$ \\ 
$p_{7}$ & $p_{8}$%
\end{tabular}%
}$ \\ \hline
$\overset{}{%
\begin{tabular}{ll}
$p_{5}$ & $p_{7}$ \\ 
$p_{6}$ & $p_{8}$%
\end{tabular}%
}$ & $\overset{}{%
\begin{tabular}{ll}
$p_{13}$ & $p_{14}$ \\ 
$p_{14}$ & $p_{16}$%
\end{tabular}%
}$%
\end{tabular}%
}
\end{equation}%
\caption{Symmetric probability table.}
\label{SymmTableEPRprobabilities}
\end{table}
By\ including the normalization constraint on each quadrant, we produce
Table (\ref{TableSymmetry}).

\begin{table}[h]
\begin{equation}
\begin{array}{c}
\text{Alice}%
\end{array}%
\begin{array}{c}
\underset{}{%
\begin{array}{c}
S_{1}%
\end{array}%
\begin{array}{c}
+1 \\ 
-1%
\end{array}%
} \\ 
\overset{}{%
\begin{array}{c}
S_{2}%
\end{array}%
\begin{array}{c}
+1 \\ 
-1%
\end{array}%
}%
\end{array}%
\overset{\overset{%
\begin{array}{c}
\text{Bob}%
\end{array}%
}{%
\begin{array}{cc}
\overset{%
\begin{array}{c}
S_{1}^{\prime }%
\end{array}%
}{%
\begin{array}{cccccc}
+1 &  &  & -1 &  & 
\end{array}%
} & \overset{%
\begin{array}{c}
S_{2}^{\prime }%
\end{array}%
}{%
\begin{array}{cccc}
& +1 &  & -1%
\end{array}%
}%
\end{array}%
}}{%
\begin{tabular}{c|c}
$\underset{}{%
\begin{tabular}{l|l}
$p_{1}$ & $p_{2}$ \\ \hline
$p_{2}$ & $1-\delta _{1}$%
\end{tabular}%
}$ & $\underset{}{%
\begin{tabular}{l|l}
$p_{5}$ & $p_{6}$ \\ \hline
$1-\delta _{2}$ & $p_{8}$%
\end{tabular}%
}$ \\ \hline
$\overset{}{%
\begin{tabular}{l|l}
$p_{5}$ & $1-\delta _{2}$ \\ \hline
$p_{6}$ & $p_{8}$%
\end{tabular}%
}$ & $%
\begin{tabular}{llll}
&  &  & 
\end{tabular}%
\overset{}{%
\begin{tabular}{l|l}
$p_{13}$ & $p_{14}$ \\ \hline
$p_{14}$ & $1-\delta _{3}$%
\end{tabular}%
}$%
\end{tabular}%
},
\end{equation}%
\caption{Symmetric normalised probability table. Here $\protect\delta %
_{1}=p_{1}+2p_{2},$ $\protect\delta _{2}=p_{5}+p_{6}+p_{8},$ and $\protect%
\delta _{3}=p_{13}+2p_{14}$.}
\label{TableSymmetry}
\end{table}
This arrangement for playing a $2\times 2$ game with $16$ joint
probabilities facilitates a transition to playing the \emph{same} game using
an Einstein-Podolsky-Rosen (EPR) type setup \cite%
{Bohm,Bell,Aspect,Peres,Cereceda}. In this setup, Alice and Bob are
spatially separated and are unable to communicate with each other. In an
individual run, both receive one half of a pair of particles originating
from a common source. In the same run of the experiment, both choose one
from two given (pure) strategies. These strategies are the two directions in
space along which spin or polarization measurements can be made. Keeping the
notation for the coins, we denote these directions to be $S_{1}$, $S_{2}$
for Alice and $S_{1}^{\prime }$, $S_{2}^{\prime }$ for Bob. Each measurement
generates $+1$\ or $-1$\ as the outcome, as it is the case with coins after
their toss in the four-coin setup. Experimental outcomes are recorded for a
large number of individual runs and payoffs are awarded according to the
directions the players choose over many runs (defining their strategies),
the matrix of the game they play, and the statistics of the measurement
outcomes.

When $p_{i}$ $(1\leq i\leq 16)$ are taken as the EPR probabilities, they of
course satisfy the normalization constraint, stating that the sums $%
\tsum\nolimits_{i=1}^{4}p_{i},$ $\tsum\nolimits_{i=5}^{8}p_{i},$ $%
\tsum\nolimits_{i=9}^{12}p_{i},$ and $\tsum\nolimits_{i=13}^{16}p_{i}$ are
all equal to $1$. EPR probabilities also satisfy other constraints imposed
by the requirements of causality, stating that Alice's outcome of $+1$ or $%
-1 $ (obtained along $S_{1}$ or $S_{2}$) is independent of whether Bob
chooses $S_{1}^{\prime }$ or $S_{2}^{\prime }$, and similarly Bob's outcome
of $+1$ or $-1$ (obtained along $S_{1}^{\prime }$ or $S_{2}^{\prime }$) is
independent of whether Alice chooses $S_{1}$ or $S_{2}$. This can be written
as 
\begin{equation}
\begin{array}{ll}
p_{1}+p_{2}=p_{5}+p_{6}, & p_{1}+p_{3}=p_{9}+p_{11}, \\ 
p_{9}+p_{10}=p_{13}+p_{14}, & p_{5}+p_{7}=p_{13}+p_{15}, \\ 
p_{3}+p_{4}=p_{7}+p_{8}, & p_{11}+p_{12}=p_{15}+p_{16}, \\ 
p_{2}+p_{4}=p_{10}+p_{12}, & p_{6}+p_{8}=p_{14}+p_{16},%
\end{array}
\label{Constraints on Ps}
\end{equation}%
which is also referred to as the \emph{causal communication constraint }\cite%
{Cereceda}. These provide two more dependencies giving $%
p_{6}=p_{1}+p_{2}-p_{5}$ and $p_{8}=p_{14}+p_{16}-p_{6}$.

Note that the causal communication constraints (\ref{Constraints on Ps}) are
set into two groups. The first group states that the probability of
obtaining a particular outcome ($+1$ or $-1$) on Alice's side of the EPR
type apparatus is independent of which one of the two measurements are
performed on Bob's side. Similarly, the second group states that the
probability of obtaining a particular outcome ($+1$ or $-1$) on Bob's side
of the EPR type apparatus is independent of which one of the two
measurements are performed on Alice's side. These constraints hold even when
Alice and Bob share an entangled state, and the non-factorizable probability
sets we consider below to construct quantum games always respect the causal
communication constraints. The causal communication constraint is also
sometimes referred to as `parameter independence', `simple locality',`signal
locality'\ or `physical locality'\ and prevents the acausal exchange of
classical information between different parts of a quantum system. \ This\
fundamental constraint is therefore retained even for the cases when the
probabilities become non-factorizable.

Substituting these relations we finally obtain the probability table for
symmetric games given by Table (\ref{TableCausalSymmetry}) in terms of the
five independent variables $p_{1},p_{2},p_{5},p_{13},p_{14}$.

\begin{table}[h]
\begin{equation}
\begin{array}{c}
\text{Alice}%
\end{array}%
\overset{%
\begin{array}{r}
\text{Bob}%
\end{array}%
}{%
\begin{tabular}{c|c}
$%
\begin{array}{c}
S_{1}%
\end{array}%
\begin{array}{c}
+1 \\ 
-1%
\end{array}%
\overset{\overset{%
\begin{array}{c}
S_{1}^{\prime }%
\end{array}%
}{%
\begin{array}{cccccc}
+1 &  &  & -1 &  & 
\end{array}%
}}{\underset{}{%
\begin{tabular}{c|c}
$p_{1}$ & $p_{2}$ \\ \hline
$p_{2}$ & $1-\delta _{1}$%
\end{tabular}%
\ \ }}$ & $\overset{\overset{%
\begin{array}{c}
S_{2}^{\prime }%
\end{array}%
}{%
\begin{array}{cccc}
+1 &  &  & -1%
\end{array}%
}}{\underset{}{%
\begin{tabular}{c|c}
$p_{5}$ & $\delta _{6}$ \\ \hline
$\delta _{5}$ & $1-\delta _{4}$%
\end{tabular}%
\ \ }}$ \\ \hline
$%
\begin{array}{c}
S_{2}%
\end{array}%
\begin{array}{c}
+1 \\ 
-1%
\end{array}%
\overset{}{%
\begin{tabular}{c|c}
$p_{5}$ & $\delta _{5}$ \\ \hline
$\delta _{6}$ & $1-\delta _{4}$%
\end{tabular}%
\ \ }$ & $\overset{}{%
\begin{tabular}{c|c}
$p_{13}$ & $p_{14}$ \\ \hline
$p_{15}$ & $1-\delta _{3}$%
\end{tabular}%
\ \ }$%
\end{tabular}%
\ \ }
\end{equation}%
\caption{Symmetric normalised causal probability table. Here $\protect\delta %
_{4}=p_{1}+p_{2}-p_{5}+p_{13}+p_{14},$ $\protect\delta %
_{5}=-p_{5}+p_{13}+p_{14},$ and $\protect\delta _{6}=p_{1}+p_{2}-p_{5}$}
\label{TableCausalSymmetry}
\end{table}

If $x,y\in \lbrack 0,1]$ are defined to be the probability to select $S_{1}$
over $S_{2}$ and $S_{1}^{\prime }$ over $S_{2}^{\prime }$ by Alice and Bob
respectively, then we have Alice's expected payoff given by 
\begin{equation}
\Pi _{A}(x,y)=xy\Pi _{A}(S_{1},S_{1}^{\prime })+x(1-y)\Pi
_{A}(S_{1},S_{2}^{\prime })+(1-x)y\Pi _{A}(S_{2},S_{1}^{\prime
})+(1-x)(1-y)\Pi _{A}(S_{2},S_{2}^{\prime }).
\end{equation}%
Bob's payoff is then obtained from above by interchanging $x$ and $y$.
Substituting Eqs.~(\ref{StrategyPayoffs}) and re-arranging we find 
\begin{eqnarray}
\Pi _{A}(x,y) &=&xy\Delta _{3}v_{3}+x(\Delta _{1}v_{1}-\Delta
_{2}v_{2})+y\{(a_{2}-a_{1})v_{1}+(a_{3}-a_{4})v_{2}\}  \notag \\
&+&a_{1}p_{13}+(a_{2}+a_{3})p_{14}+a_{4}(1-p_{13}-2p_{14}),
\end{eqnarray}%
where $\Delta _{1}=a_{3}-a_{1}$, $\Delta _{2}=a_{4}-a_{2}$ and $\Delta
_{3}=\Delta _{2}-\Delta _{1}$ with 
\begin{equation}
v_{1}=w,\text{ }v_{2}=u+v,\text{ }v_{3}=u+w,  \label{vDefn}
\end{equation}%
where $u=p_{1}-p_{5},$ $v=p_{2}-p_{14}$ and $w=p_{13}-p_{5}$. By symmetry we
also have for Bob 
\begin{eqnarray}
\Pi _{B}(x,y) &=&xy\Delta _{3}v_{3}+y(\Delta _{1}v_{1}-\Delta
_{2}v_{2})+x\{(a_{2}-a_{1})v_{1}+(a_{3}-a_{4})v_{2}\}  \notag \\
&+&a_{1}p_{13}+(a_{2}+a_{3})p_{14}+a_{4}(1-p_{13}-2p_{14}).
\end{eqnarray}%
For a NE we need to satisfy the relations 
\begin{eqnarray}
\Pi _{A}(x^{\ast },y^{\ast })-\Pi _{A}(x,y^{\ast }) &=&(x^{\ast }-x)[y^{\ast
}\Delta _{3}v_{3}+\Delta _{1}v_{1}-\Delta _{2}v_{2}]\geq 0  \notag \\
\Pi _{B}(x^{\ast },y^{\ast })-\Pi _{B}(x^{\ast },y) &=&(y^{\ast }-y)[x^{\ast
}\Delta _{3}v_{3}+\Delta _{1}v_{1}-\Delta _{2}v_{2}]\geq 0.
\label{symmetricNEConditions}
\end{eqnarray}%
We thus have the NE defined for symmetric games with the three variables $%
v_{1},v_{2},v_{3}$.

\subsection{When probabilities are factorizable}

If the probability table is factorizable then we can write 
\begin{eqnarray}
p_{1} &=&rr^{\prime },p_{2}=r(1-r^{\prime }),p_{3}=r^{\prime
}(1-r),p_{4}=(1-r)(1-r^{\prime })  \notag \\
p_{5} &=&rs^{\prime },p_{6}=r(1-s^{\prime }),p_{7}=s^{\prime
}(1-r),p_{8}=(1-r)(1-s^{\prime })  \notag \\
p_{9} &=&sr^{\prime },p_{10}=s(1-r^{\prime }),p_{11}=r^{\prime
}(1-s),p_{12}=(1-s)(1-r^{\prime })  \notag \\
p_{13} &=&ss^{\prime },p_{14}=s(1-s^{\prime }),p_{15}=s^{\prime
}(1-s),p_{16}=(1-s)(1-s^{\prime }),
\end{eqnarray}%
where $r,s,r^{\prime },s^{\prime }\in \lbrack 0,1]$ and $r=p_{1}+p_{2}$ and $%
s=p_{13}+p_{14}$. From symmetry we have $p_{2}=p_{3}$ which immediately
implies $r=r^{\prime }$ and also $s=s^{\prime }$ from $p_{14}=p_{15}$.

\begin{table}[h]
\begin{equation}
\begin{array}{c}
\text{Alice}%
\end{array}%
\overset{%
\begin{array}{r}
\text{Bob}%
\end{array}%
}{%
\begin{tabular}{c|c}
$%
\begin{array}{c}
S_{1}%
\end{array}%
\begin{array}{c}
+1 \\ 
-1%
\end{array}%
\overset{\overset{%
\begin{array}{c}
S_{1}^{\prime }%
\end{array}%
}{%
\begin{array}{cccccc}
+1 &  &  &  &  & -1%
\end{array}%
}}{\underset{}{%
\begin{tabular}{c|c}
$r^2$ & $r(1-r)$ \\ \hline
$r(1-r)$ & $(1-r)^2$%
\end{tabular}%
\ \ }}$ & $\overset{\overset{%
\begin{array}{c}
S_{2}^{\prime }%
\end{array}%
}{%
\begin{array}{cccccccc}
+1 &  &  &  &  &  &  & -1%
\end{array}%
}}{\underset{}{%
\begin{tabular}{c|c}
$rs$ & $r(1-s)$ \\ \hline
$s(1-r)$ & $(1-r)(1-s)$%
\end{tabular}%
\ \ }}$ \\ \hline
$%
\begin{array}{c}
S_{2}%
\end{array}%
\begin{array}{c}
+1 \\ 
-1%
\end{array}%
\overset{}{%
\begin{tabular}{c|c}
$sr$ & $s(1-r)$ \\ \hline
$r(1-s)$ & $(1-s)(1-r)$%
\end{tabular}%
\ \ }$ & $\overset{}{%
\begin{tabular}{c|c}
$s^2$ & $s(1-s)$ \\ \hline
$s(1-s)$ & $(1-s)^2 $%
\end{tabular}%
\ \ }$%
\end{tabular}%
\ \ }
\end{equation}%
\caption{Factorizable probabilities.}
\label{FactorizedTable}
\end{table}
So we can now find from Eqs.~(\ref{vDefn}) $v_{1}=-s(r-s)$, $%
v_{2}=(r-s)(1-s) $ and $v_{3}=(r-s)^{2}$. Substituting these results into
Eq.~(\ref{symmetricNEConditions}) gives the following conditions for the
strategy pair $(x^{\ast },y^{\ast })$ to be a NE 
\begin{eqnarray}
\Pi _{A}(x^{\ast },y^{\ast })-\Pi _{A}(x,y^{\ast }) &=&(x^{\ast
}-x)(r-s)[\Delta _{3}\{y^{\ast }r+(1-y^{\ast })s\}-\Delta _{2}]\geq 0,
\label{AliceFactNE} \\
\Pi _{B}(x^{\star },y^{\star })-\Pi _{B}(x^{\star },y) &=&(y^{\ast
}-y)(r-s)[\Delta _{3}\{x^{\ast }r+(1-x^{\ast })s\}-\Delta _{2}]\geq 0.
\label{BobFactNE}
\end{eqnarray}%
These are the defining equations for a NE when assuming symmetry and
factorizability.

For Alice, we have the payoff in the factorizable case 
\begin{eqnarray}
\Pi _{A}(x,y) &=&xy\Delta _{3}(r-s)^{2}+x(r-s)(\Delta _{3}s-\Delta
_{2})+y(r-s)(\Delta _{3}s+a_{3}-a_{4})]  \notag \\
&+&a_{4}-s(\Delta _{2}-a_{3}+a_{4})+\Delta _{3}s^{2}
\label{AlicePayoffFactorizable}
\end{eqnarray}%
and a similar expression for Bob is obtained by exchanging $x$ for $y$.

\subsection{Obtaining the classical mixed strategy game}

To achieve the classical payoff structure we see from the first term in Eq.~(%
\ref{AlicePayoffFactorizable}) that $(r-s)^{2}=1$, which requires $r=1$ and $%
s=0$ to give the payoff 
\begin{equation}
\Pi
_{A}(x,y)=a_{4}+x(a_{2}-a_{4})+y(a_{3}-a_{4})+xy(a_{1}-a_{2}-a_{3}+a_{4}),
\end{equation}%
giving the required classical bilinear payoff structure, which has
associated NE given by 
\begin{equation}
(x^{\ast }-x)[\Delta _{3}y^{\ast }-\Delta _{2}]\geq 0.
\label{AliceFactNEClassical}
\end{equation}

\subsubsection{Prisoners' Dilemma}

For the PD game we have $\Delta _{1},\Delta _{2}>0$ and hence $|\Delta
_{3}|\leq \Delta _{2}$. This makes the term in the square bracket in Eq.~(%
\ref{AliceFactNE}) and Eq.~(\ref{BobFactNE}) to be always negative, hence we
just require $r>s$ if $(x^{\ast },y^{\ast })=(0,0)$ is to exist as a NE. The
condition $r>s$ implies that the coins are basically in a heads up state,
which is obviously reasonable because if we invert the coins before the game
then we invert the NE. This shows that symmetry and factorizability along
with the condition $r>s$ will return the classical NE for the PD game.

\subsubsection{Stag Hunt}

For the SH game we have $\Delta _{3}>\Delta _{2}>0$ and $\Delta _{1}+\Delta
_{2}>0$ and $\Delta _{3}>\Delta _{1}+\Delta _{2}$. For mixed NE we require
from Eq.~(\ref{AliceFactNE}) $y^{\ast }\Delta _{3}(r-s)+\Delta _{3}s-\Delta
_{2}=0$ or 
\begin{equation}
y^{\ast }=(\Delta _{2}/\Delta _{3}-s)/(r-s).
\end{equation}%
Because, by definition, $y^{\ast }\geq 0$ then this requires $s\leq \Delta
_{2}/\Delta _{3}$ and similarly because $y^{\ast }\leq 1$ then $[\Delta
_{2}-s\Delta _{3}]/[\Delta _{3}(r-s)]\leq 1$ or $\Delta _{2}\leq \Delta
_{3}r $ or 
\begin{equation}
r\geq \Delta _{2}/\Delta _{3}.  \label{rCondition}
\end{equation}%
To create the classical mixed NE in the classical game, we define $r=\Delta
_{2}/\Delta _{3}+(1-\Delta _{2}/\Delta _{3})g$ where $g\in (0,1]$ and $%
s=\Delta _{2}/\Delta _{3}(1-h)$ where $h\in (0,1]$. This gives us $(\Delta
_{2}/\Delta _{3}-s)/(r-s)=\Delta _{2}/\Delta _{3}$ or $h\Delta _{2}/\Delta
_{3}/((1-\Delta _{2}/\Delta _{3})g+\Delta _{2}/\Delta _{3}h)=\Delta
_{2}/\Delta _{3}$ or that $g=h$. This result indicates that $r$ and $s$ are
a proportional distance from $\Delta _{2}/\Delta _{3}$. This then gives that 
\begin{equation}
s=\frac{1-r}{\Delta _{3}/\Delta _{2}-1}.  \label{sCalc}
\end{equation}%
For the other NE, $(x^{\ast },y^{\ast })=(0,0)$, if we have $y^{\ast }=0$
then we require from Eq.~(\ref{AliceFactNE}) 
\begin{equation}
\Pi _{A}(x^{\ast },y^{\ast })-\Pi _{A}(x,y^{\ast })=\frac{1}{2}(x^{\ast
}-x)(r-s)[\Delta _{3}s-\Delta _{2}]\geq 0,
\end{equation}%
and in order to return $x^{\ast }=0$ requires $s<\Delta _{2}/\Delta _{3}$.
Also, if we have $y^{\ast }=1$ then we require 
\begin{equation}
\Pi _{A}(x^{\ast },y^{\ast })-\Pi _{A}(x,y^{\ast })=\frac{1}{2}(x^{\ast
}-x)(r-s)[\Delta _{3}r-\Delta _{2}]\geq 0,
\end{equation}%
and in order to return $x^{\ast }=1$ requires $r>\Delta _{2}/\Delta _{3}$.
Hence we find three NE 
\begin{eqnarray}
(x^{\ast },y^{\ast }) &=&(0,0)  \notag \\
(x^{\ast },y^{\ast }) &=&(\Delta _{2}/\Delta _{3},\Delta _{2}/\Delta _{3}) 
\notag \\
(x^{\ast },y^{\ast }) &=&(1,1)
\end{eqnarray}%
conditional on Eq.~(\ref{rCondition}) and Eq.~(\ref{sCalc}). We know $%
0<\Delta _{2}/\Delta _{3}<1$, hence we can always find an $r$ and an $s$ to
create this particular classical game.

\subsubsection{Chicken game}

For the Chicken game we have $\Delta _{3}=-(\alpha +\beta )<0$ and $\Delta
_{2}=-\alpha <0$ and $\Delta _{1}=\beta >0$, where $\alpha ,\beta >0$. The
general condition for NE is obtained from Eq.~(\ref{AliceFactNE}) as 
\begin{equation}
\frac{1}{2}(x^{\ast }-x)(r-s)[-y^{\ast }(\alpha +\beta )(r-s)-(\alpha +\beta
)s+\alpha ]\geq 0.
\end{equation}%
We can see that $\Delta _{2}/\Delta _{3}=\alpha /(\alpha +\beta )$ hence we
will duplicate the results of the previous SH game, obtaining the correct
classical NE 
\begin{eqnarray}
(x^{\ast },y^{\ast }) &=&(1,0)  \notag \\
(x^{\ast },y^{\ast }) &=&(\alpha /(\alpha +\beta ),\alpha /(\alpha +\beta ))
\notag \\
(x^{\ast },y^{\ast }) &=&(0,1)
\end{eqnarray}%
provided $r>\alpha /(\alpha +\beta )$ and $s=(1-r)/\{(\alpha +\beta )/\alpha
-1\}$.

\subsubsection{Discussion}

We find $\alpha /(\alpha +\beta )=\Delta _{2}/\Delta
_{3}=(a_{4}-a_{2})/(a_{4}-a_{2}-a_{3}+a_{1})$, hence for the three games
studied, if we select $r$ such that $%
(a_{4}-a_{2})/(a_{4}-a_{2}-a_{3}+a_{1})<r\leq 1$ with $s$ given by Eq.~(\ref%
{sCalc}), in each case we will return the classical NE for these three games
when the probability table becomes factorizable, although not the classical
bilinear payoffs. If we also require this payoff structure, then we require
the more restrictive constraint $r=1$ and $s=0$.

\section{Extension towards non-factorizable joint probabilities}

We have shown that factorizability along with symmetry and the conditions $%
r=1$ and $s=0$ embeds the classical game within the quantum game. If we
enforce $r=p_{1}+p_{2}=1$ in the general quantum game, then we have $%
p_{3}=p_{4}=0$ by normalization, but by symmetry $p_{2}=0$, and therefore $%
p_{1}=1$, similarly for the rest of the table. Hence this condition creates
the table of factorizable probabilities with $p_{1}=p_{6}=p_{11}=p_{16}=1$
and all other probabilities zero.

However we can still create a non-factorizable set of probabilities by
inserting offset parameters from the starting position in the Table (\ref%
{FactorizedTable}). We now add extra parameters into Table \ref%
{FactorizedTable} exploiting any available degrees of freedom. For the upper
left quadrant because we are constrained by normalization, symmetry, and the
causal communication constraint, we only have available two degrees of
freedom. This is utilized with the parameters $a$ and $b$ as shown in Table (%
\ref{FactorizedNonFact}). We then continue this process and we find that we
can add up to $5$ independent parameters, $a,b,c,d,e \in \Re $ in the range $%
[-1,1]$.

\begin{table}[h]
\begin{equation}
\begin{array}{c}
\text{Alice}%
\end{array}%
\overset{%
\begin{array}{r}
\text{Bob}%
\end{array}%
}{%
\begin{tabular}{c|c}
$%
\begin{array}{c}
S_{1}%
\end{array}%
\begin{array}{c}
+1 \\ 
-1%
\end{array}%
\overset{\overset{%
\begin{array}{c}
S_{1}^{\prime }%
\end{array}%
}{%
\begin{array}{cccccc}
+1 &  &  &  &  & -1%
\end{array}%
}}{\underset{}{%
\begin{tabular}{c|c}
$r^{2}-a-2b$ & $r(1-r)+b$ \\ \hline
$r(1-r)+b$ & $(1-r)^{2}+a$%
\end{tabular}%
\ \ }}$ & $\overset{\overset{%
\begin{array}{c}
S_{2}^{\prime }%
\end{array}%
}{%
\begin{array}{cccccccc}
+1 &  &  &  &  &  &  & -1%
\end{array}%
}}{\underset{}{%
\begin{tabular}{c|c}
$rs+e$ & $r(1-s)-a-b-e$ \\ \hline
$s(1-r)+d+c-e$ & $(1-r)(1-s)+\eta $%
\end{tabular}%
\ \ }}$ \\ \hline
$%
\begin{array}{c}
S_{2}%
\end{array}%
\begin{array}{c}
+1 \\ 
-1%
\end{array}%
\overset{}{%
\begin{tabular}{c|c}
$sr+e$ & $s(1-r)+c+d-e$ \\ \hline
$r(1-s)-a-b-e$ & $(1-s)(1-r)+\eta $%
\end{tabular}%
\ \ }$ & $\overset{}{%
\begin{tabular}{c|c}
$s^{2}+c$ & $s(1-s)+d$ \\ \hline
$s(1-s)+d$ & $(1-s)^{2}-c-2d$%
\end{tabular}%
\ \ }$%
\end{tabular}%
\ .\ }
\end{equation}%
\caption{Parameterizing non-factorizability.}
\label{FactorizedNonFact}
\end{table}
In Table (\ref{FactorizedNonFact}) $\eta =a+b+e-c-d$, and $a,b,c,d,e $
chosen such that each one of the $16$ probabilities in Table (\ref%
{FactorizedNonFact}) remains in the range $[0,1]$.

From Table (\ref{FactorizedNonFact}) and Eqs.~(\ref{vDefn}) we find 
\begin{equation}
v_{1}=-s(r-s)-\epsilon _{1},\text{ }v_{2}=(1-s)(r-s)-\epsilon _{2},\text{ }%
v_{3}=(r-s)^{2}-\epsilon _{3}
\end{equation}%
and we can parameterize the NE in terms of the three parameters $\epsilon
_{1}=e-c$, $\epsilon _{2}=a+b+d+e$ and $\epsilon _{3}=a+2b-c+2e$.
Substituting $v_{1},v_{2},v_{3}$ into Eq.~(\ref{symmetricNEConditions}) we
find 
\begin{eqnarray}
&&\Pi _{A}(x^{\ast },y^{\ast })-\Pi _{A}(x,y^{\ast })  \notag \\
&=&(x^{\ast }-x)[y^{\ast }\Delta _{3}v_{3}+\Delta _{1}v_{1}-\Delta _{2}v_{2}]
\notag \\
&=&(x^{\ast }-x)(r-s)[\Delta _{3}\{y^{\ast }(r+\frac{\epsilon _{1}-\epsilon
_{3}}{r-s})+(1-y^{\ast })(s+\frac{\epsilon _{1}}{r-s})\}-\Delta _{2}(1+\frac{%
\epsilon _{1}-\epsilon _{2}}{r-s})].  \notag \\
&&  \label{NonFactorizableNEAlice}
\end{eqnarray}%
The corresponding inequality for Bob is then obtained by interchanging $x$
and $y$. This gives us the general conditions for a NE in the
non-factorizable case. Note that if we set $\epsilon _{1}=\epsilon
_{2}=\epsilon _{3}=0$ we recover our previous results in Eq.~(\ref%
{AliceFactNE}) and Eq.~(\ref{BobFactNE}). \ It is also easily shown that
this condition also implies that $a=b=c=d=e=0$ and so we will recover the
factorizable payoff relation shown in Eq.~(\ref{AlicePayoffFactorizable})

\subsection{Non-factorizable game with classical embedding}

The embedding of the classical game is obtained by taking $r=1$ and $s=0$,
which from Eq.~(\ref{NonFactorizableNEAlice}) gives the equation for NE 
\begin{eqnarray}
&&\Pi _{A}(x^{\ast },y^{\ast })-\Pi _{A}(x,y^{\ast })  \notag \\
&=&(x^{\ast }-x)[\Delta _{3}\{y^{\ast }(1-\epsilon _{3})+\epsilon
_{1}\}-\Delta _{2}(1+\epsilon _{1}-\epsilon _{2})]  \notag \\
&&\Pi _{B}(x^{\ast },y^{\ast })-\Pi _{B}(x^{\ast },y)  \notag \\
&=&(y^{\ast }-y)[\Delta _{3}\{x^{\ast }(1-\epsilon _{3})+\epsilon
_{1}\}-\Delta _{2}(1+\epsilon _{1}-\epsilon _{2})].
\label{NonFactorizableClassicalEmbedding}
\end{eqnarray}%
Thus we have obtained the general conditions for the NE for a
non-factorizable table of probabilities, which will embed the classical game
when it becomes factorizable. The payoff given by 
\begin{eqnarray}
\Pi _{A}(x,y) &=&a_{4}+c(a_{1}-a_{4})-d(2a_{4}-a_{2}-a_{3})+xy\Delta
_{3}(1-\epsilon _{3})  \notag \\
&+&x[(a_{2}-a_{4})(1-\epsilon _{2})+\epsilon
_{1}(a_{1}-a_{3})]+y[(a_{3}-a_{4})(1-\epsilon _{2})+\epsilon
_{1}(a_{1}-a_{2})]  \label{NonFactorizableNEClassicalEmbeddingPayoff}
\end{eqnarray}%
and similarly for Bob.

These produce Table \ref{FactorizedNonFactClassicalEmbedding} and we see
that we must have $a,b,c,d,e \ge 0 $.

\begin{table}[h]
\begin{equation}
\begin{array}{c}
\text{Alice}%
\end{array}%
\overset{%
\begin{array}{r}
\text{Bob}%
\end{array}%
}{%
\begin{tabular}{c|c}
$%
\begin{array}{c}
S_{1}%
\end{array}%
\begin{array}{c}
+1 \\ 
-1%
\end{array}%
\overset{\overset{%
\begin{array}{c}
S_{1}^{\prime }%
\end{array}%
}{%
\begin{array}{cccccc}
+1 &  &  &  &  & -1%
\end{array}%
}}{\underset{}{%
\begin{tabular}{c|c}
$1-a-2 b$ & $b$ \\ \hline
$b$ & $a$%
\end{tabular}%
\ \ }}$ & $\overset{\overset{%
\begin{array}{c}
S_{2}^{\prime }%
\end{array}%
}{%
\begin{array}{cccccccc}
+1 &  &  &  &  &  &  & -1%
\end{array}%
}}{\underset{}{%
\begin{tabular}{c|c}
$e$ & $1-a-b-e$ \\ \hline
$d+c-e$ & $a+b+e-c-d$%
\end{tabular}%
\ \ }}$ \\ \hline
$%
\begin{array}{c}
S_{2}%
\end{array}%
\begin{array}{c}
+1 \\ 
-1%
\end{array}%
\overset{}{%
\begin{tabular}{c|c}
$e$ & $c+d-e$ \\ \hline
$1-a-b-e$ & $a+b+e-c-d$%
\end{tabular}%
\ \ }$ & $\overset{}{%
\begin{tabular}{c|c}
$c$ & $d$ \\ \hline
$d$ & $1-c-2d$%
\end{tabular}%
\ \ }$%
\end{tabular}%
\ \ }
\end{equation}%
\caption{Non-factorizable probabilities that embed the classical game.}
\label{FactorizedNonFactClassicalEmbedding}
\end{table}

\subsection{The CHSH inequalities}

Cereceda \cite{Cereceda} finds the CHSH sum of correlations \cite{Peres} for
any set of local hidden variables satisfying the causal communication
constraint as 
\begin{equation}
\Delta =2(p_{1}+p_{4}+p_{5}+p_{8}+p_{9}+p_{12}+p_{14}+p_{15}-2).
\end{equation}%
From Table \ref{FactorizedNonFactClassicalEmbedding}, we can find 
\begin{equation}
\Delta =4(a-c+2e-1/2).
\end{equation}%
Inspecting the table of probabilities, we note that $a-c+2e-\frac{1}{2}\in
\lbrack -1,1]$, therefore, a range of possible $\Delta \in \lbrack -4,4]$
exist in agreement with the expected range \cite{Cereceda}. For example,
using $a=\frac{1}{2},b=0,c=0,d=\frac{1}{2},e=\frac{1}{2}$, we find $\Delta
=4 $. However, quantum mechanics enforces extra restrictions on the joint
probabilities considered here that can arise, namely Cirel'son's bound \cite%
{Cirelson} of $\Delta \in \lbrack -2\sqrt{2},2\sqrt{2}]$. That is, for a
physically realizable quantum game, we will have extra restriction on the
table of probabilities 
\begin{equation}
|a-c+2e-1/2|\leq 1/\sqrt{2}.  \label{quantumGameRestriction}
\end{equation}

\subsection{Quantum Prisoners' Dilemma constructed from non-factorizable
joint probabilities}

For PD, we usually take \cite{EWL} $a_{1}=3,$ $a_{2}=0,$ $a_{3}=5,$ and $%
a_{4}=1$ in matrices (\ref{A and B matrices}) and the strategy pair $%
(x^{\ast },y^{\ast })=(0,0)$ is a NE at which both players receive the
payoffs of $1$. To find if non-factorizability permits achieving $(x^{\ast
},y^{\ast })=(1,1)$ as a NE we note from Eq.~(\ref%
{NonFactorizableNEClassicalEmbedding}) that this requires 
\begin{equation}
\Delta _{3}(1-\epsilon _{3}+\epsilon _{1})-\Delta _{2}(1+\epsilon
_{1}-\epsilon _{2})\geq 0.  \label{NonFactorizableNEClassicalEmbedding11}
\end{equation}%
Here we refer to a result in Ref. \cite{Cereceda} giving a set of
non-factorizable joint probabilities that saturates Cirel'son's bound, while
maximally violating CHSH inequality. For this set we have $a=d=e=\frac{1}{8}%
(2+\sqrt{2})$ and $b=c=\frac{1}{2}-\frac{1}{8}(2+\sqrt{2})$. This results in 
$1-\epsilon _{3}+\epsilon _{1}=0$ and $1+\epsilon _{1}-\epsilon _{2}=0$, and
we have the situation of non-factorizability giving $(x^{\ast },y^{\ast
})=(1,1)$ as a NE. From Eq.~(\ref{NonFactorizableNEClassicalEmbeddingPayoff}%
) we find that the payoff for each player at this NE as 
\begin{equation}
\Pi _{A}(1,1)=\Pi _{B}(1,1)=\frac{1}{8}(18+\sqrt{2})=2.42678,
\end{equation}%
which is above the payoff of $1$ to each player at the classical NE of $%
(x^{\ast },y^{\ast })=(0,0)$ and is close to the Pareto optimum payoff of $3$
for each player.

\subsection{Stag Hunt game with non-factorizable joint probabilities}

From Eq.~(\ref{NonFactorizableClassicalEmbedding}), we now have the mixed NE
given by 
\begin{equation}
y^{\ast }=\frac{\Delta _{2}(1+\epsilon _{1}-\epsilon _{2})}{\Delta _{3}}-%
\frac{\epsilon _{1}}{1-\epsilon _{3}},
\end{equation}%
where we find $(1+\epsilon _{1}-\epsilon _{2})\in \lbrack -1,1]$ , $\epsilon
_{1}\in \lbrack -1,\frac{1}{2}]$ and $1-\epsilon _{3}\in \lbrack -\frac{1}{2}%
,1]$, so that any mixed NE we desire in the range $[0,1]$, as well as
returning to the classical NE when $\epsilon _{1}=\epsilon _{2}=\epsilon
_{3}=0$. If we desire to produce the non-classical NE of $(x^{\ast },y^{\ast
})=(0,1)$ and $(x^{\ast },y^{\ast })=(1,0)$, then from Eq.~(\ref%
{NonFactorizableClassicalEmbedding}) we have the conditions 
\begin{eqnarray}
\Pi _{A}(1,0)-\Pi _{A}(x,0) &=&(1-x)(\Delta _{3}\epsilon _{1}-\Delta
_{2}(1+\epsilon _{1}-\epsilon _{2}))\geq 0  \notag \\
\Pi _{B}(0,1)-\Pi _{B}(0,y) &=&(1-y)(\Delta _{3}\epsilon _{1}-\Delta
_{2}(1+\epsilon _{1}-\epsilon _{2}))\geq 0.  \notag \\
&&  \label{NonFactorizableNEClassicalEmbedding}
\end{eqnarray}%
We have $\Delta _{3}>\Delta _{2}>0$ for the SH game and so we can see that
if we select $\epsilon _{1}\geq 1+\epsilon _{1}-\epsilon _{2}$, then we will
have achieved this new NE. This condition gives $\epsilon _{2}\geq 1$ or $%
a+b+d+e\geq 1$. This is easily satisfied, with $b=d=1/2$ with the other
terms zero, for example.

\subsection{Chicken game with non-factorizable joint probabilities}

The Chicken game is defined with $\Delta _{2},\Delta _{3}<0$ whereas the SH
game has $\Delta _{2},\Delta _{3}>0$. \ Thus we can carry over the results
from the previous section, except that the NE will invert due to the extra
minus sign in Eq.~(\ref{NonFactorizableNEClassicalEmbedding}).

\section{Discussion}

Quantum versions of $2\times 2$ games are developed considering the
peculiarities of a set of quantum mechanical joint probabilities. The
probability sets we consider consist of normalized probabilities satisfying
causal communication and the symmetry constraints. Players are allowed
classical strategies only and their payoff relations are re-expressed in
terms of the joint probabilities. We allow a quantum game thus defined to
reduce itself to the classical mixed-strategy game when the set of joint
probabilities can be factorized in terms of the factorization parameters $r$
and $s$. Constraints on the parameters $r$ and $s$ are obtained with which
this reduction can be realized.

Non-factorizable sets of joint probabilities are introduced and appropriate
parameters $\epsilon _{1},\epsilon _{2},$ and $\epsilon _{3}$ describing
non-factorizability are identified. Quantum games are now constructed by
retaining the obtained constraints on the parameters $r$ and $s$ and
allowing non-factorizability parameters $\epsilon _{1},\epsilon _{2},$ and $%
\epsilon _{3}$ to take non-zero values. Two types of games are identified:
Firstly, with $\epsilon _{1},\epsilon _{2},\epsilon _{3}=0$, and $r=1$, $s=0$%
, we obtain the original classical mixed-strategy game along with its
bilinear payoff structure. Secondly, while enforcing $r=1$ and $s=0$, but
allowing non-factorizability parameters $\epsilon _{1},\epsilon
_{2},\epsilon _{3}$ to take non-zero values, we obtain an extension of the
classical mixed-strategy game in which the full original classical game,
along with its bilinear payoff structure, remains embedded when $\epsilon
_{1},\epsilon _{2},\epsilon _{3}=0$. We investigate PD within this setup to
find that when Cirel'son's bound is maximally saturated, a non-factorizable
and quantum game gives the NE of $(x^{\ast },y^{\ast })=(1,1)$ at which both
players' payoffs approach to their Pareto optimum value. For the SH we
observe that two new and non-classical NE of $(x^{\ast },y^{\ast })=(0,1),$ $%
(1,0)$ can be realized with non-factorizable joint probabilities. We
demonstrate that the our non-factorizable extension of the classical game
permits us to study situations that are not even physically realizable. That
is, the situations in which the corresponding CHSH inequality is violated
beyond Cirel'son's bound. We then obtain a constraint, given by Eq.~(\ref%
{quantumGameRestriction}), that defines the boundaries of what quantum
mechanics can permit for the extension of the original classical $2\times 2$
game.

Notice that the new parameters $a,b,c,d$ and $e$ are introduced in order to
give an idea of the extent of how much non-factorizable a given table of
joint probabilities is relative to the factorizable situation. So as to
obtain compact expression, the parameters $\epsilon _{1},\epsilon
_{2},\epsilon _{3}$ are then introduced, each of which depends on $a,b,c,d$
and $e$. These extra parameters are added numerically to the $r$ and $s$
parameters \textit{without} affecting the meaning of the $r$ and $s$
parameters. It turns out that $r$ and $s$ subsequently become redundant in
our quantum game as we set $r=1$ and $s=0$ in order to embed the classical
game within the quantum game.

The extension of game theory based on the considerations of quantum
mechanical joint probabilities attaining the peculiar character of being
non-factorizable has already been investigated in earlier publications \cite%
{IqbalCheon2007,IqbalCheonM,IqbalCheonAbbott,IqbalAbbott2008}. The present
manuscript's contribution consists in understanding how placing an extra
symmetry requirement on joint probabilities changes the role of
non-factorizability in the construction of quantum games. Using
probabilistic considerations only, this work explores quantum games that are
constructed using an EPR type setting. This approach gives a more accessible
perspective on the nature of quantum mechanical joint probabilities and
their potential exploitation in giving an extension to game theory.

One important benefit of this extension is the extended perspective it
provides of looking at the quantum mechanical probabilities that is able to
cover the classical factorizable, non-factorizable, and even those
situations that quantum mechanics does not allow, within a single framework.
Quantum mechanics is a probabilistic theory and this paper holds that
probabilistic considerations permit us to have a more clear vision and sense
of what quantum mechanics can achieve and what are its limits. From this
viewpoint we give an extension to game theory, while focussing on purely
probabilistic considerations. We observe that this extension is general
enough to show us the classical factorizable situations as well as the
situations that are beyond quantum mechanics.

We are motivated to have an entirely probabilistic approach towards quantum
games that encompasses classical, quantum and also those hypothetical
situations that cannot be realized quantum mechanically. We allow players
the same sets of classical strategies so that this scheme is not subjected
to Enk and Pike type argumentation. As an EPR type apparatus is used in the
playing of the two-player quantum games, their physical realization will
involve performing EPR type experiments. These experiments are agreed to
entail genuinely quantum features. The sets of joint probabilities, whose
non-factorizable property we use in constructing our quantum games, are
relevant to generalized EPR type experiments.

The potential benefits of this approach consists of developing an entirely
probabilistic understanding of multi-party strategic situations. In these
situations, quantum probabilities become crucial in achieving one or the
other outcome. The extension of game theory advocated in this paper, rather
than being imaginary, is simply a more generic framework, that allows us to
consider, in entirely probabilistic terms, also that peculiar domain which
resides beyond the bounds of quantum mechanics. We find that, within our
probabilistic approach toward quantum games, this domain becomes more easily
identifiable, along with easily recognizable classical factorizable and the
quantum mechanical non-factorizable domains.

\end{document}